\documentstyle[11pt,newpasp,twoside,epsf]{article}
\def\edcomment#1{\iffalse\marginpar{\raggedright\sl#1\/}\else\relax\fi}
\reversemarginpar

\begin{document}
\title{The DLAs contribution to the metal content of the high z Universe}
 \author{Gustavo A. Lanfranchi \& Amancio C.S. Fria\c ca}
\affil{Instituto Astron\^omico e Geof\'\i sico, USP, 
        Av. Miguel Stefano 4200, 04301-904 S\~ao Paulo, SP, Brazil}

\begin{abstract}
We investigate the evolutionary history of the Universe's metal  
content focusing on the chemical abundance of several elements 
(N, O, S, Si, Fe, Cr, Zn) taken from observational data and 
predictions from chemical evolution models. The estimated abundances  
were observed in Damped Lyman $\alpha$ Systems (DLAs) over a  
wide range of redshift (z$\sim$0.5-4.5). These data are compared to 
predictions of chemical evolution  
models. Since the nature of the DLAs is uncertain, they are represented by two
class of models: models with galactic winds describing
dwarf galaxies and with
infall representing disk galaxies. In order to
 settle constraints for star
formation timescales in
DLAs, we use the ratios [$\alpha$/Fe] and
[N/$\alpha$] in the comparison to
the predictions of the models. These ratios
in DLAs are only partially
reproduced by the disk and dwarf galaxy models
suggesting that the DLAs come
from a variety of morphological types of
galaxies and not only one. They
also imply a typically long timescale for the
star formation in these
systems.
\end{abstract}

\section{Introduction}

Much effort has been devoted to the study of chemical abundances in the
DLAs in the last few years (Lu et al. 1996,
Centuri\'on et al 2000, Prochaska $\&$ Wolfe 1999, Ellison et al. 2001,
 Molaro et al. 2001). Most of these works is new
observational data with subsequent analysis, 
which is  based in comparison to known patterns of abundance and 
abundance ratios, mainly the ones observed in our Galaxy. These studies
have already revealed some important characteristics of the DLAs which
gives us some clues about their
nature: their metallicity, given generally by [Zn/H], lies in the 
range -2 to 0 (Vladilo 1998, Lauroesch et al. 1996)  
suggesting different epochs for DLAs formation or different star  
formation histories; there is no sign of evolution in the metallicity 
of DLAs from N(HI)-weighted Fe abundance (Prochaska
$\&$ Wolfe 1999, Prochaska $\&$ Wolfe 2000), altought the metallicity 
content does decrease with redshift (Savaglio 2000); the [Zn/Fe] 
ratio is mostly above solar indicating the presence of dust depletion 
in these systems (Pettini et al. 1997); the [$\alpha$/Fe] pattern 
observed in DLAs roughly resembles that of metal-poor stars of our 
Galaxy (Pettini et al. 2000, Ellison et al. 2001), even though, there 
are some systems with values similar to those predictions by chemical 
evolution models for dwarf galaxies (Centurion et al. 2000, Molaro et 
al. 2001).

Among these works, however, just in a few of them chemical evolution models
are used in the comparison to the observational data (Prantzos $\&$ Bossier
2000, Matteucci et al. 1997). The predictions of the chemical evolution 
models are compared to absolute abundances (Prantzos $\&$ Boissier 2000) and
abundance ratios (Matteucci et al. 1997). The abundance ratios are better
suited for chemical evolution studies because, unlike absolute values,  
they do not depend heavily on model assumptions, but mainly on the stellar  
nucleosynthesis and on the adopted IMF. The [$\alpha$/Fe] and 
[N/$\alpha$]  
ratios can be used to settle constraints for star formation timescales  
due to the different timescales for the formation of these
elements. While the $\alpha$ elements are produced mainly in type II  
Supernovae (SNe II) in short timescales, the iron peak ones and N are
produced in SNe Ia and  in intermediate massive stars
(IMS) in a larger timescale. There is a controversy
about the observed pattern of the [$\alpha$/Fe] ratio in DLAs: 
while
some authors detected patterns similar to that of the metal-poor stars  
of our Galaxy (Pettini et al. 2000, Ellison et al. 2001), others found 
values that resemble the chemical evolution of dwarf galaxies  
(Centurion et al. 2000, Molaro et al. 2001). This discrepancy may be  
associated with dust depletion. 

We use, in this 
work, one-zone chemical evolution models to represent the DLAs, which
predictions are compared to [$\alpha$/Fe] and [N/$\alpha$] ratios. The
abundance of Zn, Fe and Si are corrected for dust depletion and robust
statistical methods are used to establish the trends of the data. In this way,
we try to make a complete study of the DLAs chemical  
evolution embracing the most important aspects that might affect these 
kind of analysis. 

\section{The Chemical Evolution Models}

Due to uncertainties about the nature of DLAs, two types of  
one-zone model are used in a first study about chemical  
enrichment in these systems. A model with infall of pristine gas is used to  
represent the disk systems and another one with galactic  
winds for dwarf galaxies.  
 
\subsection{The disk model} 
 
The disk model belongs to the class of the models with one  
infall episode of pristine gas. We used the basic equations for chemical
evolution, with a  star formation rate (SFR) similar to the one used by
Matteucci $\&$ Fran\c cois (1989), and a Salpeter IMF. The
normalisation $\tilde\nu$
of the star formation law is made with the solar
position at 8 kpc and a galaxy age of 13 Gyr. The stellar yields
for
intermediate mass stars (IMS) come
from Renzini $\&$ Voli, 1981 (hereafter RV81), models with $\alpha _c$ = 0 and
1.5. The yields from van den Hoek $\&$ Groenewegen  
(1997) for the IMS were also tested but the models with  
these yields produce less nitrogen than both
$\alpha_C$ = 0 and 1.5 models of RV81 and consequently are not applied in
this DLAs study. The infall rate
decreases with time and the infall timescale
and superficial mass density
profile vary with the galactic radius.
	 
We run several models with different values of  
r: 2, 4, 8, 14, 18 kpc. 
 
\subsection{The dwarf galaxy model} 
 
The dwarf galaxy model belongs to the class of models  
with galactic winds. The majority of  
the models used here are similar to the classic ones, 
in which a wind is established in a time $t_w$ when the  
thermal energy of the gas exceeds the potential  
energy. 
	 
A galaxy with 1 Gyr is represented by the models  
here applied, but there is also a model which represents a  
galaxy with $10^7$ yr. The galaxy is inside a dark  
halo with a mass three times greater than the initial galaxy's  
baryonic mass. The star formation is continuous with a SFR proportional to
the present mass of gas and characterised by the specific star formation  
rate $\nu$ (the inverse of the star formation timescale),  
which varies in each model from 0.1 to 19 $Gyr^{-1}$.  
It is adopted a Salpeter IMF between 0.1 and 100 M$_{\odot}$.  
	 
Besides the classic models with wind, an {\it early wind}  
model (Fria\c ca 2000) is applied. Fria\c ca (2000) proposed  
differential winds, due to SNeII, in dwarf galaxies in order to  
explain the [$\alpha$ /Fe] below solar in the  
intercluster medium.  

\section{The Data} 
 
The observational data about the DLAs were collected  
from the literature and include 
chemical abundance (represented by [X/H]= $log(X/H)_{obs}$ -
$log(X/H)_{\odot}$) of a variety of elements ranging from N
to Zn, observed in the
redshift interval $0.6 < z < 4.4$ (for more
details and references see Lanfranchi $\&$ Fria\c ca 2001, submited).

\section{Analysis of the data}  
 
The observational data are analysed with robust statistical methods and
compared to predictions of the chemical evolution models. In the
statistical analysis we followed the approach described by Chiappini et al.
(1999). We obtain, with the summary statistics of Cleveland $\&$ Kleiner
(1975), three lines that summarise the trend of the data: MM (midmean), LSMM
(lower semi-midmean) and
USMM (upper semi-midmean).
The final plots represent locally weighted regression (lowess) smoothes of the
summary lines. 

The trend lines and the points distribution are then compared to the
predictions of the models described above focusing the [$\alpha$/Fe] and
[N/$\alpha$] ratios. Due to the [Zn/Fe] generally above solar in our sample,
what may be an indication of the presence of dust in the DLAs, we corrected
the abundances of Zn, Cr, Fe and Si by dust depletion. The correction is based
in the depletion pattern observed in our Galaxy together with the observed
[Zn/Fe] in each system analysed (see Lanfranchi $\&$ Fria\c ca 2001). The
[Si/Fe] corrected for dust is then used as a tracer of the [$\alpha$/Fe] ratio
in the comparison of the data to the predictions of the models, while [N/S]
is used to represent the [N/$\alpha$] ratio. In this way, we obtain dust free
(or almost free) tools to be compared with the models.

\section{Application of the one-zone models to DLAs} 
 
\subsection{Disk galaxy model} 
 
The disk models predictions for [$\alpha$/Fe] and [N/$\alpha$] ratio, using  
the RV81 yields with $\alpha _C$ = 1.5, for different radius, r = 2, 4, 8,   
14, 18 kpc compared to 
the observed values in DLAs are shown on the figure 1.

The observed [N/S]  
ratios with absolute values are centralised mainly between  
the lines for r = 4 and 14 kpc, with a large number of systems  
around the r = 8 kpc line, as indicated by the statistical trend lines. The  
lines trend are also similar to the models predicted behaviour: the  
ratio increases with [$\alpha$/H] rapidly in low metallicities 
and then becomes flatter. 

The comparison between the observed [N/$\alpha$] in DLAs and   
the disk model predictions indicates that the observed values   
of this ratio represent a long timescale for the star   
formation in these systems for regions around the solar   
neighbourhood ($4 < r < 14\; kpc$).  

\begin{figure}  
\plottwo{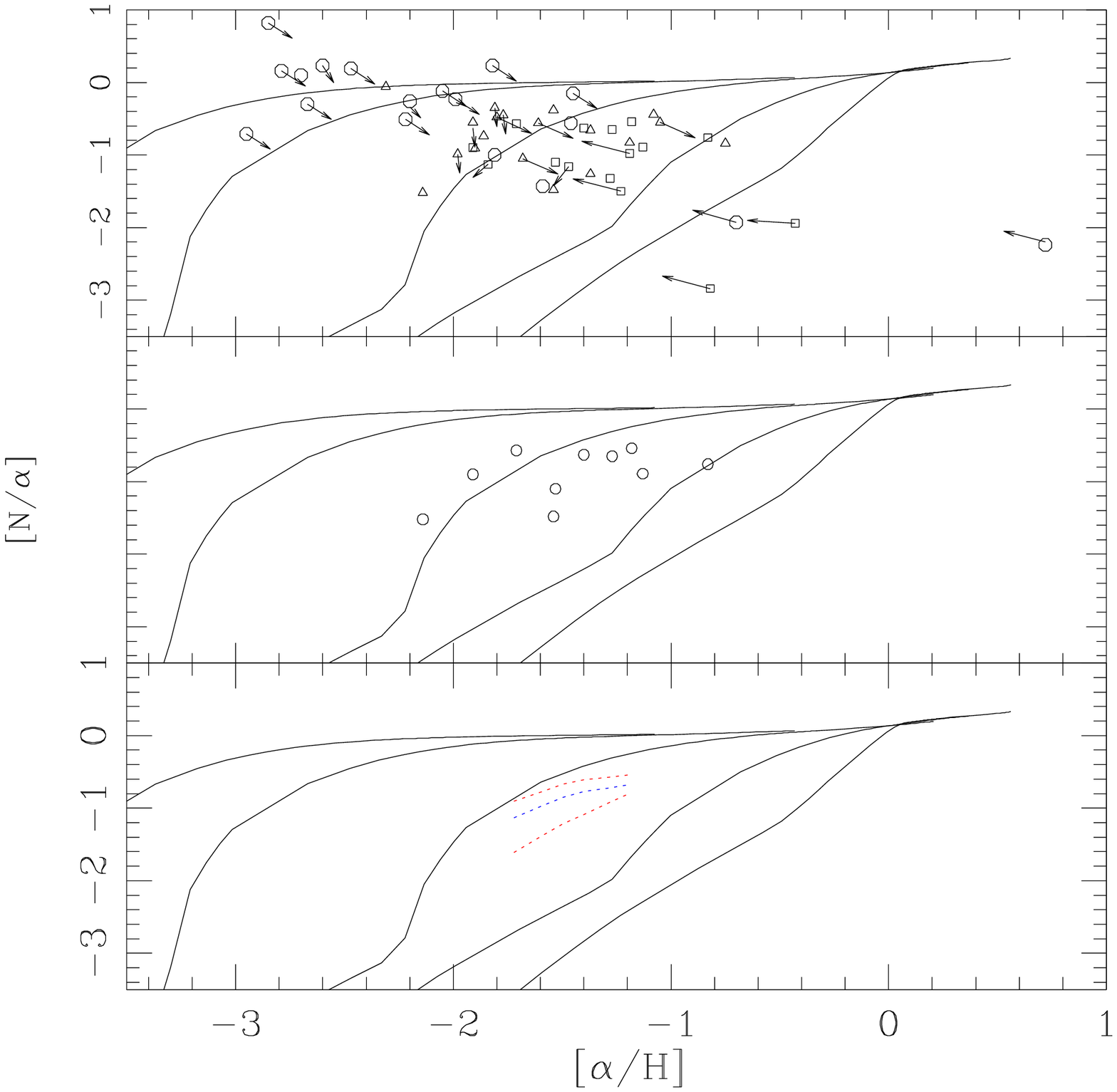}{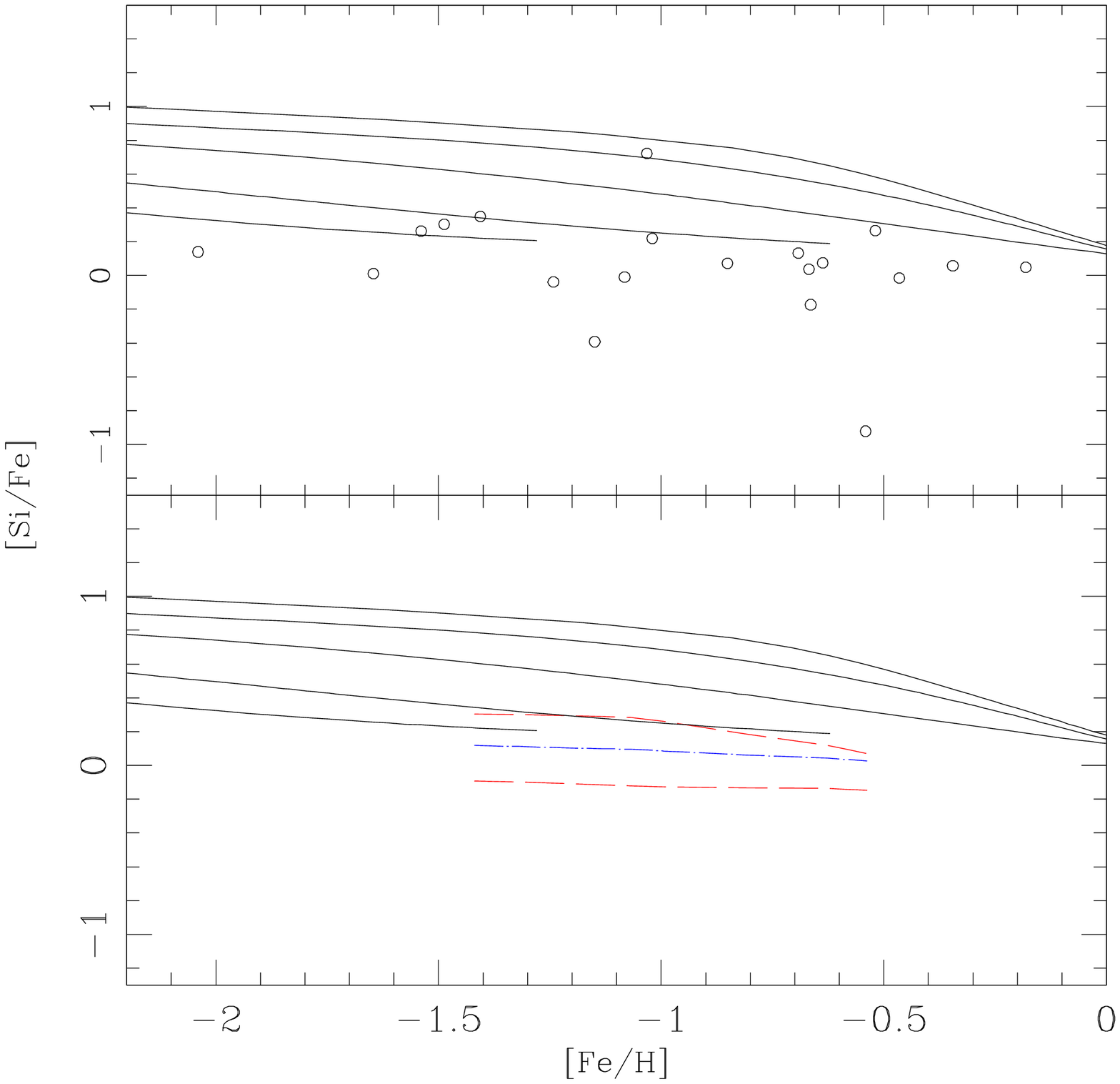}  
\caption{[N/$\alpha$] (left) and [$\alpha$/Fe] (right) observed in DLAs
compared to the disk models. The $\alpha$
elements abundance are represented  on the top left panel 
by oxygen (open circles), sulphur (squares) and silicon (triangles).   
The [N/S] ratio free of dust is plotted in the middle left panel. The
straight lines represent the grid of models  
for r = 2, 4, 8, 14, 18 kpc (right  to left and top to bottom).  
The statistical lines are the MM (long-dash - dot), USMM and LSMM (long dash)
.}   
\end{figure}

There is a discrepancy, however, when the corrected [Si/Fe] ratio is used in
the comparison to the models. The region of the models with $r \approx$ 14 -
18 kpc fits reasonably well the majority of the systems. The same behaviour
is found for the statistical lines. They exhibit a trend similar to the
models predictions and are located in the region around
$\approx\;14-18\;kpc$. Consequently, the comparison indicates that the
observed values correspond to external regions ($r > 14\;kpc$) of disk
systems, implying a long star formation timescale. A long and continuous
star formation allows the iron peak group elements to be produced, giving
rise to low values of [Si/Fe].

\subsection{The dwarf galaxy model} 
  
On the figure 2, the comparison between the predictions of the dwarf
galaxy models with different specific star formation rates and  the
[$\alpha$/Fe] and [N/$\alpha$] ratios are shown.

The predictions for [N/$\alpha$] are compared  
to the observed ratios both in DLAs and dwarf galaxies. The $\alpha$  
elements abundance is determined only by oxygen (circles) in dwarf galaxies  
while in DLAs sulphur (squares) and silicon (triangules) are also used to
estimate it. The statistical lines are the same as on previous figure.

\begin{figure}  
\plottwo{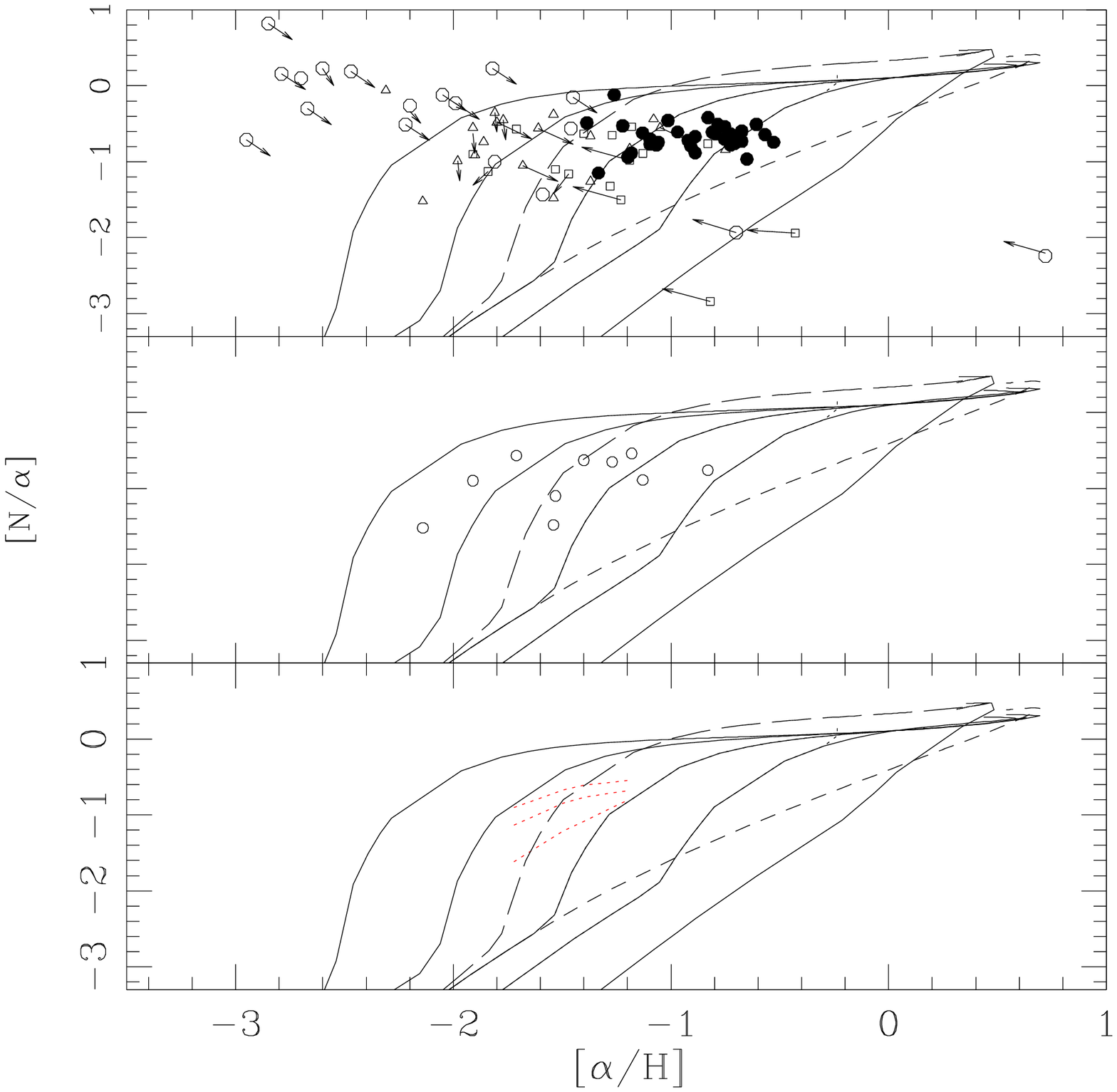}{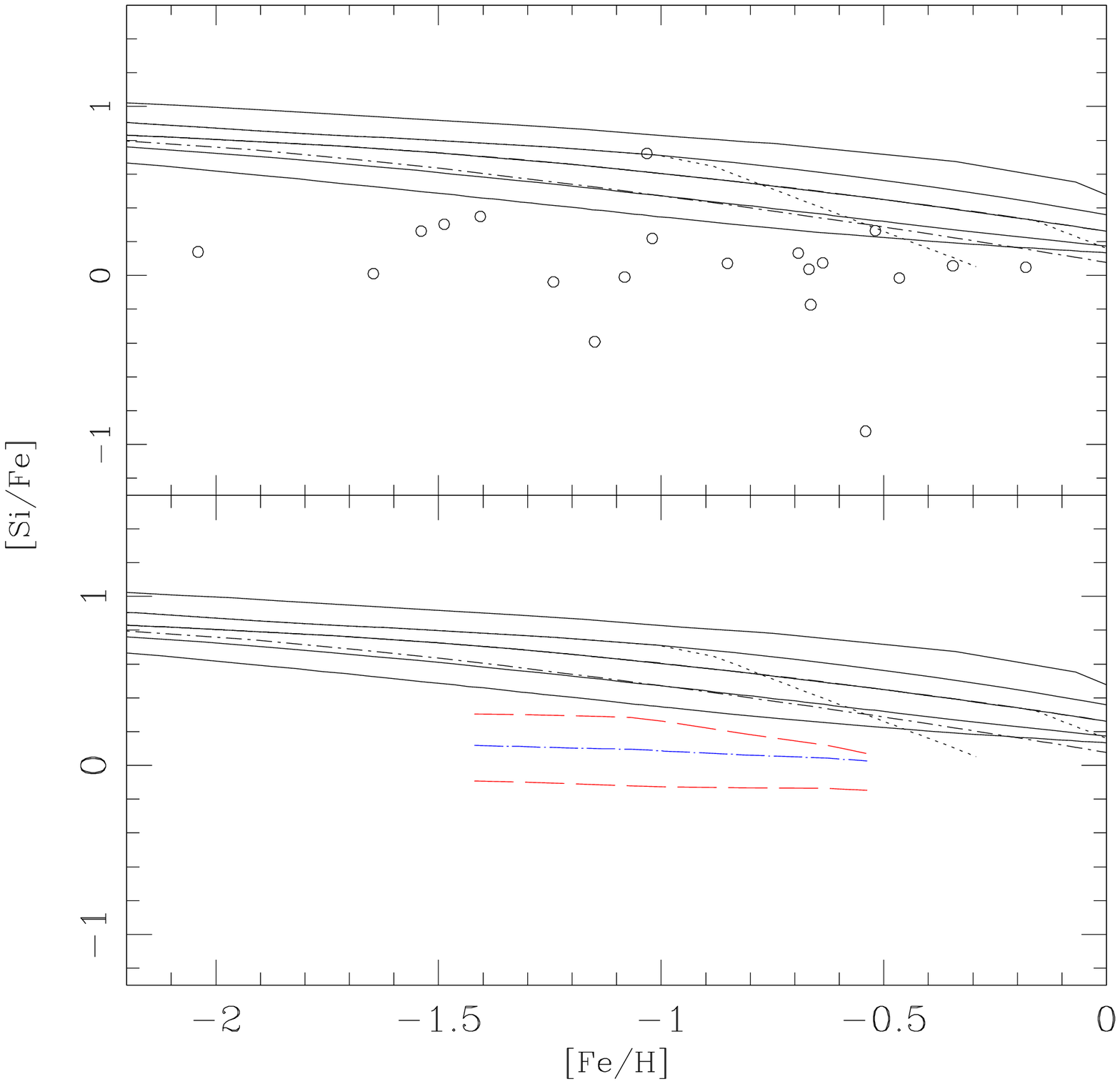}  
\caption{[N/$\alpha$](left) and [$\alpha$/Fe](right) observed in DLAs and dwarf
galaxies compared to dwarf galaxy models. The
[N/S] ratio is plotted in the middle left panel. The straight  
lines represent the grid of models with RV81 with $\nu$ = 0.1, 0.3,   
1, 3, 19 $Gyr^{-1}$ (left to right and bottom to top), the long dashed line   
a model with $\nu$ = 1.0 $Gyr^{-1}$ and {\it early wind}, and the   
dashed line a model with $\nu$ = 3.0 $Gyr^{-1}$ and   
$M\;=\;10^{7}\;M_{\odot}$. The statistical lines are the same as   
on the figure 1.}  
\end{figure}

The observed [N/$\alpha$] in DLAs show a great dispersion, larger 
than that found for Galactic stars in the same range of metallicity 
and for dwarf galaxies. The different values of N/$\alpha$  
for a given $\alpha$/H can be explained with the difference 
between the formation timescale of N and $\alpha$ elements.  
This fact can be seen in models with diverse values of $\nu$  
that go from almost quiescent star formation  
($\nu\;=\;0.1\;Gyr^{-1}$) to an extreme rapid one  
($\nu\;=\;19\;Gyr^{-1}$ for systems with  
$10^9 M_{\odot}$). Most dwarf galaxies seem to require a specific star  
formation ($\nu$) between 0.3 and 3 $Gyr^{-1}$, while the DLAs seem 
to need lower $\nu$: $0.1 \leq \nu \leq 1$ $Gyr^{-1}$ for classic wind models
and $\nu \approx$ 1 $Gyr^{-1}$ for models with {\it early wind}. The models
with galactic winds reproduce all dwarf galaxy points but are unable to match
some DLAs that fall totally out of the region delimited by these models. 
 
In the case of the dwarf galaxy models too, there is a difference when the
[Si/Fe] ratios is used in the comparison to the models. While for the [N/S]
ratio, most of the observed values are fitted by models with
$0.3 \leq \nu \leq 1\;Gyr^{-1}$, most systems and the statistical lines lie
below the [$\alpha$/Fe] predicted by any model. Only the upper semi-midmean
(USMM - dotted line) goes near the model with $\nu$ = 0.1 $Gyr^{-1}$. This fact
implies a specific star formation $\nu \leq 0.1\;Gyr^{-1}$, which represents a
long star formation timescale.

\section{Conclusions}

The comparison between the one-zone chemical evolution models for dwarf and
disk galaxies and the observational data indicates that these
models are able to reproduce only partially the [N/$\alpha$] and [$\alpha$/Fe]
observed in DLAs. This fact suggests that these systems may represent various
types of galaxies and not a single one. The observed abundance ratios also
require long star formation timescale in both models. The discrepancy between
the comparisons using [N/$\alpha$] and [$\alpha$/Fe] might be related to an
underestimated correction for Si.


\begin{references}
 
\reference 
Centuri\'on M., Bonifacio P., Molaro P. \& 
Vladilo G. 2000, ApJ, 536, 540 
 
\reference 
Chiappini C., Matteucci F., Beers T.C.  \& Nomoto K. 1999, ApJ, 515, 
226 
 
\reference 
Cleveland W.S. \& Kleiner B. 1975, {\it Technometrics}, 17, 447 
 
\reference 
Ellison S.L., Lewis G.F., Pettini M., Sargent W.L.W., Chaffee F.H. \& 
Irwin M.J. 1999, astro-ph 9903063 
 
\reference 
Fria\c ca A.C.S. 2000, in {\it Dwarf Galaxies and Cosmology}, Proceedings 
of the XXXIII Recontres de Moriond, Y.T. TRINH, C. BALKOWSKI, CAYETTEV, 
J.T.T. V\^AN, Paris, Editions Fronti\`eres, p.312      
  
\reference 
Lanfranchi G.A. \& Fria\c ca A.C.S. 2001, submited to MNRAS

\reference 
Lauroesch J.T., Truran J.W., Welty D.E. \& 
York D.G. 1996, PASP, 108, 641 
 
\reference 
Lu L., Sargent W.L.W., Barlow T.A., Churchill C.W. \&  
Vogt S. 1996, ApJS,  107, 475 

 
\reference 
Matteucci F., Molaro P. \& Vladilo G. 1997, AAp, 321, 45
 
\reference 
Matteucci F., \& Fran\c cois P., 1989, MNRAS, 239, 885 
 
\reference 
Molaro P., Bonifacio P., Centuri\'on M., Vladilo G., D'Odorico S., Levshakov
S.A. 2001, astro-ph 
  
\reference 
Pettini M., Ellison S.L., Steidel C.C., Shapley A. E. \& Bowen D.V. 
2000,  ApJ, 532, 65 
 
\reference 
Pettini M., King D.L., Smith L.J. \& 
Hunstead R.W.,  1997, ApJ,  478, 536 
 
\reference 
Prantzos N. \& Boissier S. 2000, MNRAS, 315, 82 

\reference 
Prochaska J.X., Naumov S.O., Carney B.W., McWilliam A. \& Wolfe A.M.
2000, astro-ph 0008075 
 
\reference 
Prochaska J.X. \&  Wolfe A.M. 1999, ApJS, 121, 369 

\reference 
Prochaska J.X. \&  Wolfe A.M., 1997, ApJ,  487, 73
 
\reference 
Renzini A. \& Voli M. 1981, A\&A, 94, 175 

 
\reference 
Savaglio S., Panagia N. \& Stiavelli M. 1999, astro-ph 9912112

\reference 
Van Den Hoek L.B. \& Groenewegen M.A.T. 1997, AASS, 123, 305 

 
\end{references}
\end{document}